\documentclass{aa}  
\usepackage{graphicx}
\usepackage{txfonts}
\usepackage{natbib}[author, year]
\usepackage{lscape}
\usepackage{soulutf8}
\usepackage[utf8]{inputenc}
\usepackage{amsmath}

\usepackage[markup=underlined]{changes}
\begin{document}

   \title{A search for new symbiotic stars in the Milky Way: }
   \subtitle{Using machine learning techniques applied to photometric databases}

   \author{V. Contreras Rojas
          \inst{1}
          \and
          M. Jaque Arancibia
          \inst{1}
          \and
          C.E. Ferreira Lopes
          \inst{2, 3}
          \and
          N. Monsalves
          \inst{3, 1}
          \and
          R. Angeloni
          \inst{5}
          \and
         Luna, G. J. M
         \inst{6, 7}
         \and V. Marels
         \inst{1}
         \and D. Concha
         \inst{2}
         \and Nuñez, N. E
         \inst{7, 8, 9}
         \and C. Saffe 
         \inst{7, 8, 9}
         \and M. Flores
         \inst{7, 8, 9}
          }

   \institute{Departamento de Astronomía, Universidad de La Serena, Raul Bitran 1720256, La Serena, Chile \\
    \email{valentina.contrerasr@userena.cl}
   \and
    Instituto de Astronomía y Ciencias Planetarias, Universidad de Atacama, Copayapu 485, Copiapó, Chile
    \and
    Millennium Institute of Astrophysics (MAS), Nuncio Monseñor Sotero Sanz 100, Of. 104, Providencia, Santiago, Chile
    \and
    Department of Physics, University of Rome Tor Vergata, via della Ricerca Scientifica 1, 00133, Rome, Italy
    \and
    Gemini Observatory/NSF’s NOIRLab, Casilla 603, La Serena, Chile
    \and
    Universidad Nacional de Hurlingham (UNAHUR), Secretaría de Investigación, Av. Gdor. Vergara 2222, Villa Tesei, Buenos Aires, Argentina
    \and
    Consejo Nacional de Investigaciones Científicas y Técnicas (CONICET)
    \and
    Facultad de Ciencias Exactas, Físicas yNaturales, Universidad Nacional de San Juan, Av. Ignacio de la Roza 590 (O), Complejo Universitario “Islas Malvinas”, Rivadavia, J5402DCS, San Juan, Argentina
    \and
    Instituto de Ciencias Astronómicas, de la Tierra y del Espacio (ICATE-CONICET), C.C 467, 5400 San Juan, Argentina
     }

   \date{Received July 15, 2025; accepted February 9, 2026}

    \abstract
    {Symbiotic stars are interacting binary systems composed of a red giant transferring material to a hot compact star, typically a white dwarf. These systems are crucial for studying stellar evolution, accretion processes, mass transfer, and a variety of complex astrophysical phenomena. However, there is a significant discrepancy between the number of confirmed symbiotic stars (\(\sim300\)) and the estimated population in the Milky Way (\(1.2 \times 10^3 - 1.5 \times 10^4\)), suggesting that a large fraction remains undetected.}
    {To address this issue, we propose the identification of new symbiotic stars through the application of machine learning techniques. Our approach combines multi-band photometric data from Gaia DR3, 2MASS, and WISE, together with parallax measurements and the pseudo-equivalent width of H\({\alpha}\), to effectively distinguish symbiotic candidates from other stellar populations.}
    {We trained a Random Forest model using a sample of 166 confirmed S-type symbiotic stars and a control sample of 1600 non-symbiotic stars. To mitigate class imbalance and improve classification performance, we applied the Synthetic Minority Oversampling Technique (SMOTE). The model achieved an \(F_1\)-score of 89\% for the symbiotic class.}
    {We applied our model to a catalog of approximately 2.5 million stars selected based on photometric colors consistent with those of S-type symbiotic stars. From this sample, 990 candidates were identified with a classification probability of at least 70\%. To refine the selection, we applied statistically and physically motivated cuts based on effective temperature, surface gravity, metallicity and complemented by SkyMapper photometry. This process yielded 12 high-confidence candidates, characterized by cool temperatures, low surface gravities, solar-like metallicity, \(H{\alpha}\) emission, luminosities ranging from moderate to high, and UV excesses consistent with the properties of S-type symbiotic systems.}
    {To evaluate the model's performance, we applied it to a validation set of  symbiotic stars recently confirmed in the literature, recovering 92.3\% of them. This result supports the effectiveness and generalizability of our classification approach.}

\maketitle
%

\section{Introduction}
\label{Introduction}

Symbiotic stars (hereafter SySts) are long-period interactive binaries composed of an evolved giant star that transfers mass to a hot compact companion, typically a white dwarf (WD), though in rare cases a neutron star (NS) may be present \citep[see, e.g.,][]{2012Mikolajewka,2019Munari}. These systems are surrounded by a circumbinary nebula formed by colliding stellar winds, photoionization, and collisions \citep{1997Muerset, 2005Kenny, 2009AKilpio, 2017Mukai}. The ongoing interaction between the three components emit across the entire electromagnetic spectrum, from radio wavelengths to X-rays \citep{2013Luna,2021ADickey}. Each spectral region is dominated by distinct physical components: the cool giant produces absorption features, while the ionized nebula gives rise to prominent emission lines. Furthermore, in approximately 50\% of cases, SySts exhibit emission in the O\,\textsc{vi} lines, a unique spectroscopic signature of these systems \citep{2000Belczynki}. Owing to their composite nature, SySts serve as valuable astrophysical laboratories for the study of a wide array of phenomena, including mass accretion, stellar winds, jet formation, thermonuclear outbursts, and stellar evolution \citep{2003Sokoloski,2016Muckai}. Moreover, they have been proposed as potential progenitors of Type Ia supernovae \citep{1992Munari,2010DiStefano,2010Wang}.

According to the evolutionary stage of the cool giant component, symbiotic systems are broadly classified into two subtypes: S-type (stellar) and D-type (dusty) systems \citep{1995MedinaTanco}. The S-type are characterized by their infrared emission that are dominated by the stellar photosphere of the giant. These systems typically contain red giants with effective temperatures between 3\,500 to 4\,000 K, corresponding to spectral types M, K, or occasionally G \citep{1975Webster,2003Mikolajewka}. Their spectral energy distribution (SED) peaks between 1.0 and 1.1 $\mu$m, consistent with photospheric emission from M3–M6 giants \citep{1995Ivison,2007Gromadzki}. The cool components are generally located on the red giant branch (RGB) or the asymptotic giant branch (AGB), and display photometric variability due to ellipsoidal modulation, eclipses, or eruptive events \citep{1999Murset}. In contrast, D-type stars are more evolved, often show Mira pulsation and exhibit significant warm dust emission, which shifts the SED to 2-2.5 $\mu$m\citep[700-1000K e.g, ][]{1982Allen, 2019Chen}. They sometimes also exhibit two shells with distinct temperatures \citep[e.g.][]{2010Angeloni}. In this case, the spectra is more similar to that of a planetary nebula than to a giant star. A small group of D-type SySts hosting G- or K-type giants shows even colder dust components. These systems are designated as D’-type and are characterized by peak SEDs at 20-30 $\mu$m. In general the SySts population in the Milky Way are dominated by the S-type (aprox ~80\%), while the D and D' only represent the 15\% and 3\% respectively.

While the cool component dominates the near-infrared and red optical spectra, the occurrence of symbiotic systems in the shorter wavelengths (UV, optical, and X-rays) is primarily determined by the hot component. Depending on whether or not the dwarf is undergoing stable hydrogen shell burning, SySts systems are classified into shell-burning or accretion-only systems \citep{2003Sokoloski, 2019Munari}. In shell burning, the hot component radiates near the Eddington limit, driving strong photoionization and producing high ionization lines, e.g., He~\textsc{ii}, Fe~\textsc{vii}, and O~\textsc{vi} \citep{1991Murset}. These systems are bright in the UV bands and often exhibit supersoft X-ray luminosities \citep{2013Luna}. On the other hand, accretion-only systems lack continuous nuclear burning and are significantly fainter, with weak or absent emission lines and low UV and X-ray luminosities. Their discovery usually relies on high-energy observations, as in the case of SU~Lyn \citep{2016Muckai}, whose optical spectrum mimics that of a normal red giant star. Recent studies have highlighted the potential of multi-wavelength screening techniques to discover these low-accretion systems. For example, \citet{2024Xu} identified several nearby candidates using UV and X-ray data combined with Gaia DR3, suggesting that such accretion-only systems might be more common than previously believed.

Theoretical estimates of the Galactic population of SySts vary significantly, by two to three orders of magnitude, when compared with observational data. The earliest estimate placed the population at $\sim4\times10^3$ \citep{1986Kenyon}, followed by a substantial increase to $\sim3\times10^5$ by \citet{1992Munari}, later revised to $\sim3.3\times10^4$ by \citet{1993Kenyon, 2003Magrini} proposed a value of $\sim4\times10^5$, assuming that 0.5\% of RG and AGB stars are in binary systems with WDs. The population synthesis models by \citet{2006Lu} suggest a range between $1.2\times10^3$. More recently \citet{2025Laverseiler} estimated lower and upper limits for the SySts population by combining empirical and binary synthesis models, setting the range between 800 and 1\,400 as a minimum and up to $(53 \pm 6) \times 10^{4}$ as a maximum.

To help close this gap, several authors have utilized large databases as tools to identify new candidates. For instance, \citet{2008Corradi} developed the first systematic search for SySts using narrowband photometric surveys, identifying stars with H${\alpha}$ excess emission detected in \citet[IPHAS][]{2005Drew} and infrared excesses based on 2MASS photometry, reporting three new discoveries, and highlighting the challenge of distinguishing SySts from their photometric mimics such as T-Tauri and Be stars. \citet{2014Rodriguez-Flores} reported 14 new SySts using optical photometry from IPHAS+, successfully confirming 5 new stars, while \citet{2015Li} identified the first halo SySts by cross-matching known systems with LAMOST DR7 spectra and added a new D-type star found in catalogs with over 4 million spectra. Similarly, \citet{2019Akrasa} conducted a targeted census of O~\textsc{vi} Raman scatterers, discovering 72 new candidate systems, though no new confirmed stars were reported.

Of course, the discovery and confirmation of a few hundred new symbiotics would place meaningful lower limits on theoretical predictions. However, the relatively low confirmation rates highlight the limitations of traditional identification methods based on individual optical spectroscopic observations  which would have failed in cases such as the case of SU~Lyn, whose symbiotic nature stood up in the high energy regime. As astronomical databases continue to grow in both size and dimensionality, manual discovery of rare objects like SySts becomes increasingly unfeasible. In this context, machine learning (ML) has emerged as a powerful approach for scalable and automated classification across high-dimensional parameter spaces. ML techniques excel at detecting subtle and consistent patterns in complex, multidimensional data, thereby improving the identification of the most promising candidates. Recent studies have begun adopting ML approaches: \citet{2019Akras,2021Akras} used decision tree classifiers to distinguish SySts from H$\alpha$-rich mimics; \citet{2023Jia} trained models on infrared photometry (2MASS and WISE cited below) to classify LAMOST sources, confirming two new SySts; and \citet{2025Ball} developed a Gaia-only pipeline integrating photometry, astrometry, and compressed XP spectra to identify over 1\,600 new candidates.

Despite the growing application of ML in identifying rare astrophysical objects, recent reviews \citep{2025Merc}, highlight persistent structural limitations in current pipelines. First, there is a noticeable lack of standardized training datasets: many models are trained on heterogeneous photometric catalogs without common calibration or physical priors. Second, typical ML classifiers often depend on narrow photometric bands (e.g., IR only), which limits their ability to generalize across different SySt subtypes. For example in \citet{2023Jia} present more than 10 thousands candidates using 2MASS and WISE photometry but only confirm 2. This highlights the need for a more physically informed ML approach, in which constraints from domain knowledge are incorporated to more effectively narrow the candidate-selection space.

In this work, we develop a supervised ML framework to identify new candidates for S-type SySts. Our decision to focus exclusively on S-type systems is motivated by physical, methodological, and computational considerations. In addition to constituting the majority of the known Galactic population, these systems exhibit more uniform photometric properties, which facilitate their characterization and distinction from other stellar types. 

Our approach leverages a combination of astrometric and photometric data from Gaia, 2MASS, and WISE, constrained within an observationally defined parameter space and specifically tailored to the characteristics of S-type SySts systems. The main goals of this work are to identify previously unknown S-type candidates and to assess the reliability of previously proposed systems using classification models trained with characterized color indices. The structure of the paper is as follows: Section~\ref{seccion2} describes the dataset construction and selected features. Section~\ref{seccion3} describes the ML methodology. Section~\ref{seccion4} presents model performance metrics. Section~\ref{seccion5} applies the trained model to the full dataset and highlights the most promising candidates. Section~\ref{seccion6} discusses how our results compare with previous approaches, and Section~\ref{seccion7} concludes with the implications of our findings.

\section{Data}
\label{seccion2}

For this study, we adopted as a reference the New Online Database of Symbiotic Variables \citep[hereafter Merc's catalogue,][]{2019Merc}\footnote{\url{https://sirrah.troja.mff.cuni.cz/~merc/nodsv/}, accessed in April 2024}, which constitutes the most comprehensive compilation to date of spectroscopically confirmed symbiotic systems, both in the Galaxy and in nearby extragalactic environments. In particular, this catalogue contain information about 1\,190 sources in the Milky Way, of which 284 are confirmed (203 S-type and 39 D-type), 690 are suspected, 78 are possible, 46 are likely and 155 are misclassified. For the definition of each category, we refer the reader to \citet{2022Merc}.

This catalog provides uniform and referenced data on observational parameters (e.g., positions and multiband photometry), as well as physical characteristics, including orbital elements, component properties and subclassifications. In addition to infrared measurements, the database integrates multiwavelength information from X-ray missions such as ROSAT and Swift \citep[respectively]{2020Evans, 2016Boller}, along with astrometric and photometric data from the Gaia mission. For the purposes of this work, we restricted our sample to confirmed Galactic systems to select the photometric databases used for training ML model.      

\begin{figure}[h!]
\centering
\includegraphics[width=1.0\linewidth]{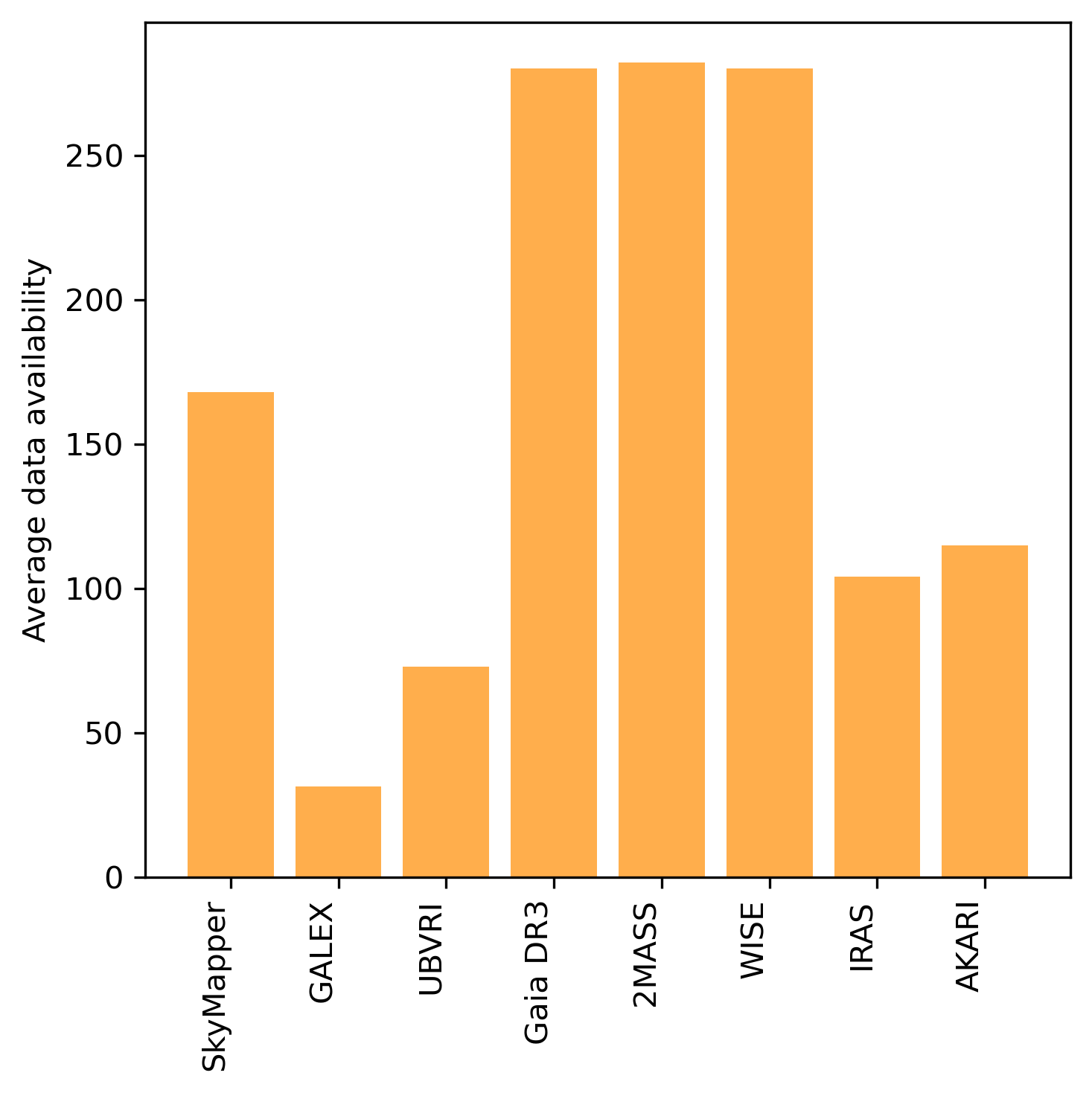}
\caption{Average number of valid measurements per photometric group across the sample. Each bar represents the mean number of available (non-missing) values for the bands belonging to a given instrument or spectral range. }
\label{fig:cantidaddedatos}
\end{figure}

The selection was guided by the completeness of photometric data available for each object, as shown in Fig.~\ref{fig:cantidaddedatos}. Among the databases considered, the ones offering the most complete coverage were Gaia DR3 \citep{2016Gaia,2023Gaiadr3}, the Two Micron All Sky Survey \citep[2MASS;][]{20062MASS}, and the Wide-field Infrared Survey Explorer \citep[WISE;][]{2010WISE}. While all four WISE bands were initially retrieved, only W1 (3.4 $\mu$m, $\Delta\lambda$ = 0.66 $\mu$m) and W2(4.6 $\mu$m, $\Delta\lambda$ = 1.04 $\mu$m) were retained for this study, as they provide an optimal trade-off between sensitivity and spatial resolution, which is essential for the ML model developed. At these last wavelengths, the infrared emission of SySts is typically dominated by the Rayleigh-Jeans tail of the giant continuum, modulated by the absorption features such as CO and H$_2$O bands. Therefore, W1 and W2 trace the photospheric emission of the cool giant, allowing for an effective separation between the S-type, characterized by a stellar continuum with minimal dust contribution, and D-type systems.

It is also important to note that fundamental stellar parameters from Gaia, such as effective temperature (\rm T$_{eff}$), surface gravity (log g), and metallicity ([Fe/H]), were excluded from the training set, since these quantities are available for only about 50\% of confirmed SySts. Given the limited number of objects with complete measurements, we chose not to apply data-imputation techniques, as this would not be statistically robust.

Then, photometric colors were obtained by combining bands from the same study to minimize the efects of intrinsic variability. Although the observations are not strictly contemporaneous, this does not significantly affect the derived colors: the typical timescales of variability in SySts, such as pulsations and orbital modulations, are larger than the time separations between bands in surveys such as Gaia, 2MASS and WISE. Therefore, while the intrinsic variability remains a relevant factor, the impact of the color measurement is expected to be limited in the context of population level classification. 

\begin{figure*}[h!]
    \centering
    \includegraphics[width=0.4\linewidth]{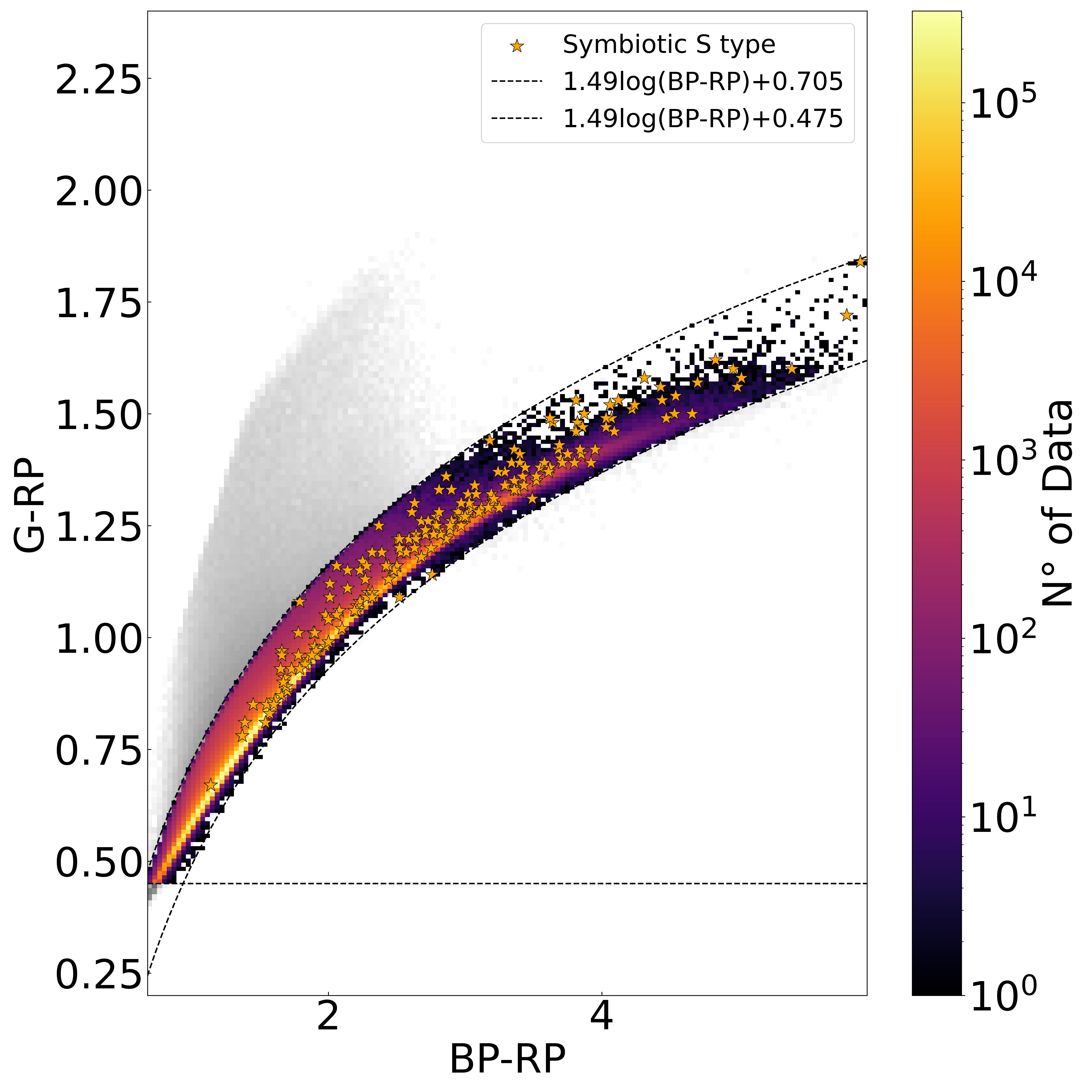}%
    \includegraphics[width=0.4\linewidth]{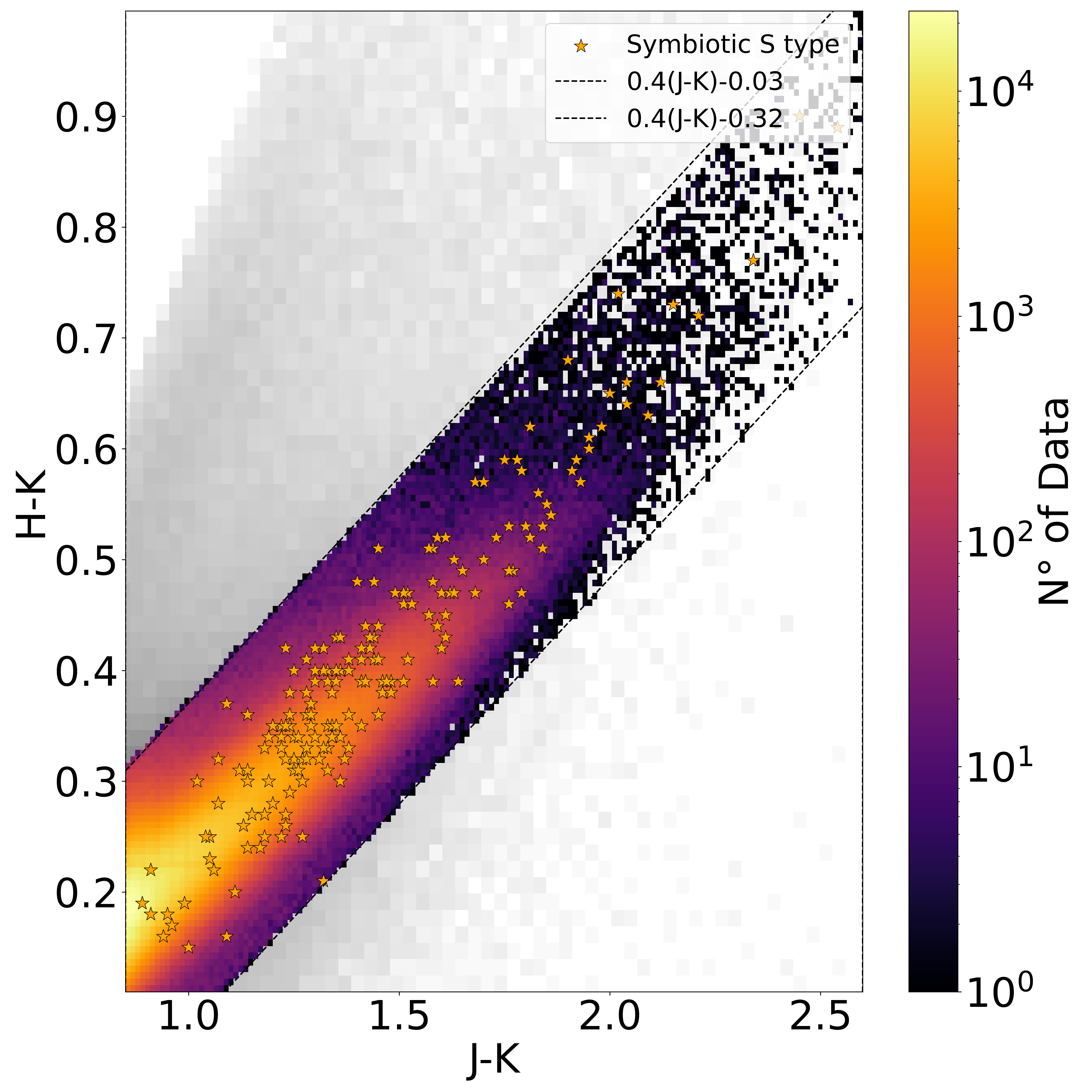}
    \caption{Color–color selection process. 
    \textbf{Left:} 2D histogram of approximately 17\,988\,392 sources obtained from four ADQL queries to the Gaia services and a WISE crossmatch. The first selection cut was applied in the Gaia color–color diagram using a linear regression in logarithmic scale, resulting in the removal of 143\,434 sources. Confirmed SySts are marked as orange stars. 
    \textbf{Right:} 2D histogram of over four million sources selected from the 2MASS color–color diagram. The color bar indicates the number of sources per bin in logarithmic scale for both panels.}
    \label{fig:colorcuts}
\end{figure*}

In this study, we limited our analysis to S-type SySts. This decision is based on the distinct photometric behavior of the different subclasses. D and D$'$ type systems possess extended dusty envelopes that significantly affect their infrared emission, especially, beyond 2 $\mu$m. These dusty components contribute to strong mid-infrared excess, which shifts their colors away from the photospheric sequence defined by the dusty-free systems. As a result, these objects occupy a broader and more dispersed region in the color-color diagrams shown in Fig. \ref{fig:colorcuts}, making it difficult to define a consistent photometric locus suitable for supervised classification.  

From this reference sample, we determined the minimum and maximum values for each color index (see Table~\ref{tab:ADQLquery}), defining a multidimensional color space representative of S-type SySts. These bounds were subsequently used to retrieve photometrically similar sources from the Gaia DR3 extended dataset. Furthermore, we adopted the observed range of pseudo-equivalent width H${\alpha}$ (${\rm EW}{\rm H\alpha}$), obtained from the Gaia DR3 astrophysical\_parameters table. These values are calculated from low-resolution XP spectra evaluated between 646 and 670 nm, and span from 0.69~\AA{} to $-$18.49~\AA{}(negative values indicate emission). Parallax values between 0 and 5.29 mas were also used as an empirical constraint to restrict the search space. These criteria were designed to favor the identification of sources with properties consistent with those of confirmed S-type SySts.

The photometric, spectroscopic, and astrometric limits defined above are critical to ensuring the quality and reliability of the training dataset used for the ML model. The Merc$'$s catalog includes a wide variety of SySts candidates and confirmed systems, compiled from heterogeneous sources, which may differ in classification criteria and observational coverage. Incorporating all available data without extensive validation could introduce inconsistencies and biases that impair model performance. Therefore, we limited our selection to confirmed S-type systems with well-characterized photometric, spectroscopic, and astrometric measurements. This approach prioritizes uniformity and robustness over sample size, allowing the model to learn from a representative and reliable dataset that reflects properties of S-type SySts.

\begin{table}[h!]
    \centering
    \caption{Summary of color limits in the ADQL query.}
    \begin{tabular}{c c c c c c}
    \hline
        Color & Min & Max & Color & Min & Max   \\ \hline
        G-BP & -4.19 & 0.10 & J-H & 0.22 & 2.99 \\
        G-RP & 0.29  & 2.34 & J-K & 0.46 & 5.33 \\
        BP-RP & 0.22  & 5.95  & H-K & 0.11 & 2.34 \\ \hline
    \end{tabular}
    \label{tab:ADQLquery}
\end{table}

After this, we performed an ADQL query using Gaia DR3 services to identify sources that matched the characteristic color ranges of S-type SySts (see Table~\ref{tab:ADQLquery}). To ensure the reliability and relevance of the selected data, several additional criteria were applied. First, we restricted the sample to sources with magnitudes in the $G$ band brighter than magnitude 16, a range in which Gaia provides higher precision astrometric and astrophysical parameters according to \citet{2019Anders}. Second, we selected only those sources with parallax measurements and ${\rm EW}{\rm H\alpha}$ values available and within the range described above. To further ensure high-confidence distance estimates, we required a parallax signal-to-noise ratio greater than 10, a threshold commonly adopted in Gaia-based studies \citep[e.g.,][]{2021Lindegren}. Finally, we required all selected sources to include a reliable counterpart in the 2MASS catalogue, using the ESA-provided cross-comparison table gaiadr3.tmass\_psc\_xsc\_best\_neighbor, to ensure consistent photometric information across all surveys.

To optimize the information without overloading the Gaia search services, the ADQL query was divided into four sky regions, each spanning a 45$^o$ declination range and covering the full 24 hours of right ascension. A simplified version of this query can be found in Appendix~\ref{app:adql}. The query yielded approximately 18 million records, representing only 1\% of the Gaia DR3 data. These results were compared to the WISE catalog using the TOPCAT tool \citep{Topcat}, with a tolerance of two arcsec. To ensure that no source is repeated in this catalog, an internal crossmatch of the catalog was performed with a tolerance of 5 arcsec, from which 1 star was excluded. This catalog, with 17\,988\,392 sources, is represented in the right panel of Fig.~\ref{fig:colorcuts} by the gray background.

With this, we start the data curation process. First, we excluded from this catalog the known SySts and restricted the W1$-$W2 color to $-$2.79 and 1.93, which corresponds to the same color range of S-type SySts. Subsequently, we constructed color-color diagrams using Gaia photometry, where it was observed that this type of stars is concentrated in a specific region of the plane. To define it, a linear regression was used with the vertical axis in logarithmic scale from which the equation was obtained:

\begin{equation}
    G-RP = 1.49{}log(BP-RP) +0.59
\end{equation}

From equation (1), a lower and upper limit were defined, which is represented in right panel of 
Fig.~\ref{fig:colorcuts} by the dashed lines in both panels using a 1$\sigma$ difference defined by visual inspection ($\sigma = 0.24 $). This same procedure was applied to the 2MASS color plane using the selected data seen in the left panel of Fig.~\ref{fig:colorcuts}. The linear regression results in an equation (2):

\begin{equation} 
    H-K_s = 0.4(J-K_s) -0.175 
\end{equation}

In this process, equations for the upper and lower bounds were defined using a  2$\sigma$ envelope ($\sigma$ = 0.15) by visual inspection. These equations are described at the top of right panels of Fig. \ref{fig:colorcuts}. After applying the color–color selection criteria described in the same figure, the filtered catalog retained approximately four million sources. Subsequently, we applied the data-cleaning procedure described by Monsalves et al. (under review), which consists of excluding the 1\% of outliers in the distribution of relevant features, photometric colors, ${\rm EW}{\rm H\alpha}$ The analytical expressions for the boundaries are shown at the top of the left panel in Fig.~\ref{fig:colorcuts}. 

This procedure further reduced the sample to approximately 2\,540\,539 objects photometrically similar to SySts. Applying the same selection and cleaning criteria to the initial set of S-type SySts from the Merc$'$s catalog reduced the sample from approximately 203 to 166 stars. This careful filtering ensures the homogeneity and reliability of the reference sample used to train the ML model. Importantly, the photometric data used in this study are neither dereddened not corrected for interstellar extinction. However, since the most distant sources in the sample are located at approximately 6 kpc, the effect of extinction is expected to be modest and not significantly affect the classification based on observed colors. Furthermore, this approach aligns with the intended application of the model to observed survey data, which are also uncorrected.

\section{Searching for SySts: ML approach} 
\label{seccion3}

\subsection{Train, Test and Validation Sets}

To build the training and test sets, we crossmatched the point source subcatalog with the SIMBAD database \citep{Simbad}, using the position with a tolerance of 2 arcsec and considering the $\texttt{main\_type}$ field. This yielded $\sim$304\,230 sources with known classifications, including various types of variable stars (e.g., Mira, Cepheids, LPVs), evolved stars (e.g., AGB, RGB, red supergiants, Wolf–Rayet), young stellar objects (e.g., T-Tauri, Herbig Ae/Be), and other peculiar Galactic sources. Many of these classes are known photometric mimics of S-type SySts in color-color space \citep{2008Corradi, 2019Akras}.

To define the negative class for the binary classifier, we first removed sources with ambiguous or uncertain types (e.g., LPV\_Candidate or LPV*), reducing the initial sample by about half. From the remaining labeled sources, we constructed the training subset by progressively subsampling the negative while enforcing statistical consistency with the parent distribution. Specifically, for all features used in the classifier (Gaia, 2MASS and WISE colors, parallax, and H$\alpha$ indices), we performed two-sample Kolmogorov–Smirnov tests \citep{1933Kolmogorov, 1948Smirnov} comparing each candidate subsample against the full negative population. Subsamples were iteratively reduced until every feature satisfied a p-value threshold of $>0.05$, ensuring that their marginal distributions remained statistically indistinguishable from those of the complete set (see Appendix~\ref{app:ks_tests} for the full list of KS statistics and p-values). The final subset of 1\,600 sources consists primarily of objects labeled as Star (1\,580), along with a small number of EclBin (12), RSG (2), YSO (2), and one each of FarIR and IR sources, thus preserving the heterogeneity of the negative class.

\begin{figure*}[h!]
    \centering
    \includegraphics[width=0.9\linewidth]{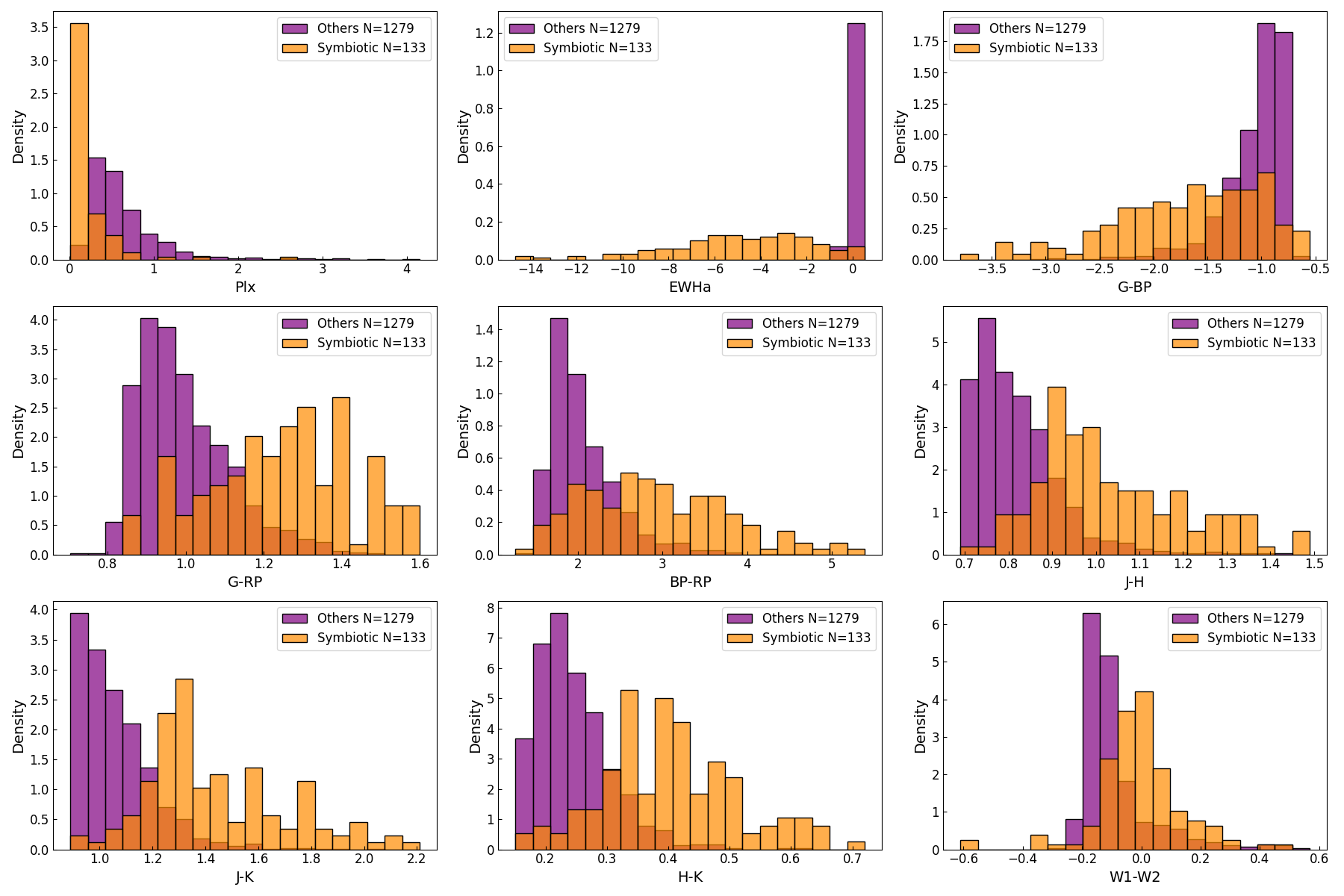}
    \caption{Distribution of training set characteristics for the positive class (S-type SySts) and the negative class ('other').}
    \label{fig:distribution}
\end{figure*}

To this negative class, we added 166 confirmed S-type SySts from Merc$'$s catalogue, which constitute the positive class. The full dataset was split using an 80:20 ratio, resulting in 133 S-type SySts and 1\,279 negative-class objects for training, and 33 SySts and 320 negative-class objects for testing, respectively. All preprocessing and data balancing steps described below were applied exclusively to the training subset, after the 80:20 split. The test partition remained untouched throughout training, calibration, and evaluation, ensuring a leakage-free workflow. The distribution of both classes is shown in Fig.~\ref{fig:distribution}.

In addition of this samples, we took an independent validation sample composed of recently confirmed symbiotic S-type stars. These sources are not included in the Merc$'$s catalog, as they were identified more recently. However, their symbiotic nature has been confirmed spectroscopically, providing a reliable and unbiased benchmark for evaluating model performance in real, never before observed cases. This validation sample builds on the work of \citet{2024Lucy}, who proposed a novel photometric selection technique using SkyMapper \citep{2011Skymapper} photometry, specifically in the $u$, $v$, and $g$ bands. Their method focused on identifying outliers in $u$-band photometry, particularly those exhibiting significant excess ultraviolet and infrared signatures consistent with S-type SySts. This approach effectively highlights systems with anomalous photometric behavior, often associated with symbiotic interactions. From their list, we used the 12 spectroscopically confirmed sources, ensuring that they were compatible with the previously defined parameter space for the classifier. By ensuring compatibility with the original feature distributions, this selection enables a robust and meaningful assessment of the model's generalization capability. Notably, none of these sources were included in the training or testing phases. Their inclusion as an external validation set allows us to test the classifier's predictive accuracy under realistic conditions, using a independently identified systems that are representative of the target class, but without training bias.

As a complement, a second validation set was incorporated, consisting of photometric impostors included in the Merc$'$s catalog as misclassifications. These correspond to objects that were initially reported in the literature as possible SySts but were later reclassified after spectroscopic confirmation. The original set comprises 155 sources distributed across the Milky Way; however, only 90 meet the consistency criteria within the adopted parameter space. These sources encompass a wide variety of stellar types that are often photometrically confused with SySts, such as variable stars, massive stars and supergiants, planetary nebulae, cataclysmic variables, subdwarfs, and young stellar objects. The inclusion of this group allows us to explore the limits of photometric separability between genuine SySts and their main contaminants that have been observationally identified.

\subsection{Data Balancing Algorithm}

The training sub-dataset consisted of 1\,279 sources from the negative class and only 133 confirmed S-type SySts from the positive class, resulting in a significant imbalance with a ratio close to 90:10. To address this problem, we applied the Synthetic Minority Oversampling Technique \citep[SMOTE,][]{2002Chawla}, an algorithm widely used to address class imbalance in classification problems. SMOTE creates new synthetic examples of the minority class, generating interpolations between each object and its nearest neighbors in the feature space defined for the classifier. This approach expands the minority class while preserving the structure of the original data.

SMOTE was applied to the training subset after the 80:20 split, while the test and validation sets remained untouched during the analysis. This resulted in a balanced dataset. The effectiveness of SMOTE in astronomical applications has been demonstrated in several recent works addressing highly unbalanced datasets. Some examples are \citet[][]{2004Chen, 2020Hosenie, 2022Maravelias, 2023Avdeeva}.

\subsection{Random Forest Model}

We employ a Random Forest (RF) binary classifier \citep{2001Breiman}, implemented using the Python library \citep[\texttt{imblearn};][]{2017imblearn}, to distinguish candidates of S-type SySts from photometric mimics. RF is a supervised ensemble learning method that builds a collection of decision trees, each trained on bootstrap samples of the original dataset. At each node, a random subset of features is considered, introducing diversity among the trees and reducing their correlation. This ensemble strategy lowers the variance of the model and mitigates the overfitting typically observed in individual decision trees, enhancing its generalization capacity in complex classification tasks involving correlated astrophysical parameters.

The model was trained using nine input features that combine photometric and astrometric information: seven color indices derived from Gaia, 2MASS, and WISE, the (${\rm EWH\alpha}$), and parallax. Hyperparameter tuning was performed via exhaustive grid search with 5-fold cross-validation using GridSearchCV \citep{sklearn_api}, applied to the training subset, leading to a final configuration with 800 trees, a minimum of ten samples to split a node, two sample per leaf, and $\log2$ of the total number of features considered at each split and  10 maximum depth was imposed, allowing the trees to grow sufficiently deep to capture complex, non-linear patterns. This setup enables the model to effectively learn the intricate feature interactions and distributions characteristic of astrophysical data, yielding a robust and reliable classifier for the identification of S-type SySts candidates within large and heterogeneous samples.

\section{Results}
\label{seccion4}

To evaluate the performance of the RF Classifier, we employed four widely recognized metrics: precision, recall, F1-score, and PR-AUC \citep{2011DPowers}. These metrics provide a comprehensive assessment of the model’s predictive capabilities, focusing on different aspects of classification performance.

Precision quantifies the ability to minimize false positives, recall measures the model's capacity to identify true positives, and the F1-score balances precision and recall to offer a single performance measure. Finally, the PR-AUC summarizes the classifier’s performance across varying classification thresholds.

Table~\ref{tab:scores} summarizes the RF model’s classification performance, evaluated via repeated stratified 5-fold cross-validation (3 repetitions). The classifier achieves near-perfect scores for the negative class (“Others”) and robust results for the symbiotic class, with 93\% precision, 85\% recall, and an F1-score of 89\%. The slightly lower recall reflects the inherent class imbalance, but overall the model reliably distinguishes SySts. Furthermore, the analysis of error rates as a function of the classification threshold shows that the false positive rate decreases from 0.6\% to 0.3\% when increasing the decision boundary from 0.5 to 0.7, while the false negative rate rises from 15.2\% to 27.3\%. These trends indicate that stricter thresholds substantially reduce contamination at the cost of a moderate loss in completeness.

\begin{table}[h!]
    \centering
    \caption{Summary of the performance scores per class for the RF Classifier applied to test sample with their respective  standard deviation.}
    \begin{tabular}{c c c} \hline
    Metric & Others & Symbiotic \\ \hline
    Precision & 0.99 $\pm$ 0.01 & 0.93 $\pm$ 0.08\\
    Recall & 0.99 $\pm$ 0.01 & 0.85 $\pm$  0.01 \\
    F1-Score & 0.99 $\pm$ 0.01 &  0.89 $\pm$ 0.04 \\
    PR-AUC Score & 0.94 $\pm$ 0.04  & 0.95 $\pm$  0.04 \\ \hline
    \end{tabular}
    \label{tab:scores}
\end{table}

\begin{figure}[h!]
    \centering
    \includegraphics[width=0.85\linewidth]{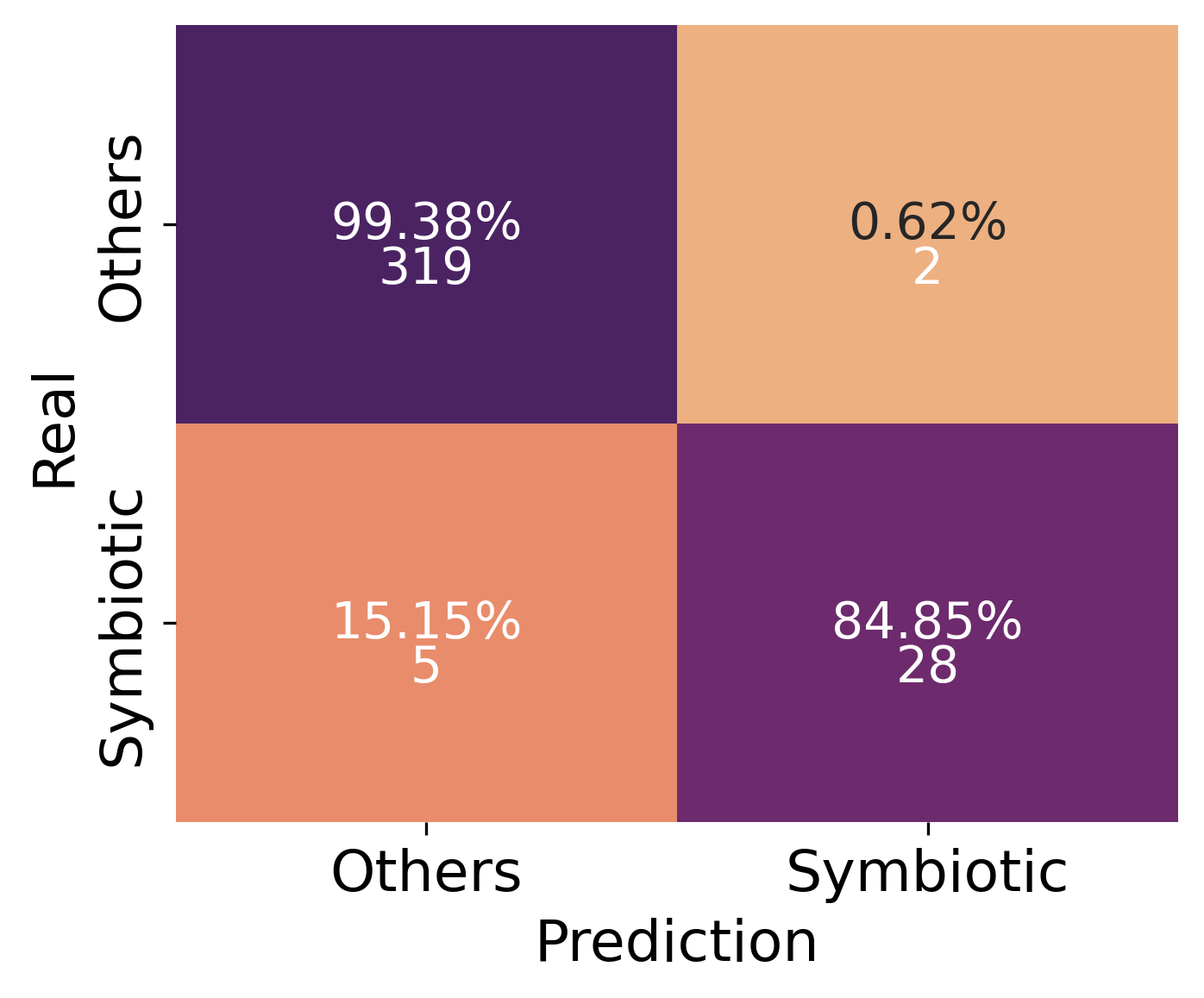}
    \caption{Confusion matrix for the testing set using SMOTE + RF, incorporating photometric colors, parallax, and ${\rm EW}{\rm H\alpha}$. The x-axis represents the predicted class (predicted label), while the y-axis denotes the actual class (true label). Each cell shows the percentage relative to its respective class on the first line, followed by the corresponding number of stars on the second line.}
    \label{fig:confussionmatrix}
\end{figure}

To reduce FP contamination and increase confidence in candidate identification, we enforce a stricter classification threshold of 70\%. This setting prioritizes precision over recall, which is especially relevant for underrepresented classes like SySts and allows for a compromise between data quantity and data quality. The confusion matrix in Fig.~\ref{fig:confussionmatrix} illustrates the classification performance on the test set, composed by a 20\% of the training catalogs, which consists of 321 objects from the negative class and 33 from the positive class (SySts).

The model correctly identifies 319 out of 321 non-symbiotic sources based on their stellar parameter estimates. Only two FP were found, corresponding to an eclipsing binary and a young stellar object. Additionally, five FN were identified: four of them are accretion-only SySts, one of which contains a NS companion, and the remaining case is a shell-burning system characterized by strong Raman emission. The results highlight the conservative nature of classifier, which prioritized a low FP rate, an essential criterion for effective spectroscopic follow-up, particularity in the context of rare-objects search. Unlike previous studies that present a large amount of candidates (around hundreds of thousands), often with limited reliability, our approach emphasize precision over quantities. In practice, the utility of a large sample of candidates is significantly reduced if it includes a substantial number of FP, as spectroscopic confirmation of all these candidates is highly unlikely. A model that minimized the probability of contamination is therefore more valuable, even at the expense of completeness. The small but highly reliable sample produced by our classifier offers a more efficient path toward confirmation, maximizing the scientific return of follow-up observations. The low number of FP demonstrates the model's robustness in separating genuine symbiotic systems from photometric mimics, while the few FN reflect the intrinsic heterogeneity and observational diversity within the SySts class.

\begin{figure}[h!]
    \centering
    \includegraphics[width=1.0\linewidth]{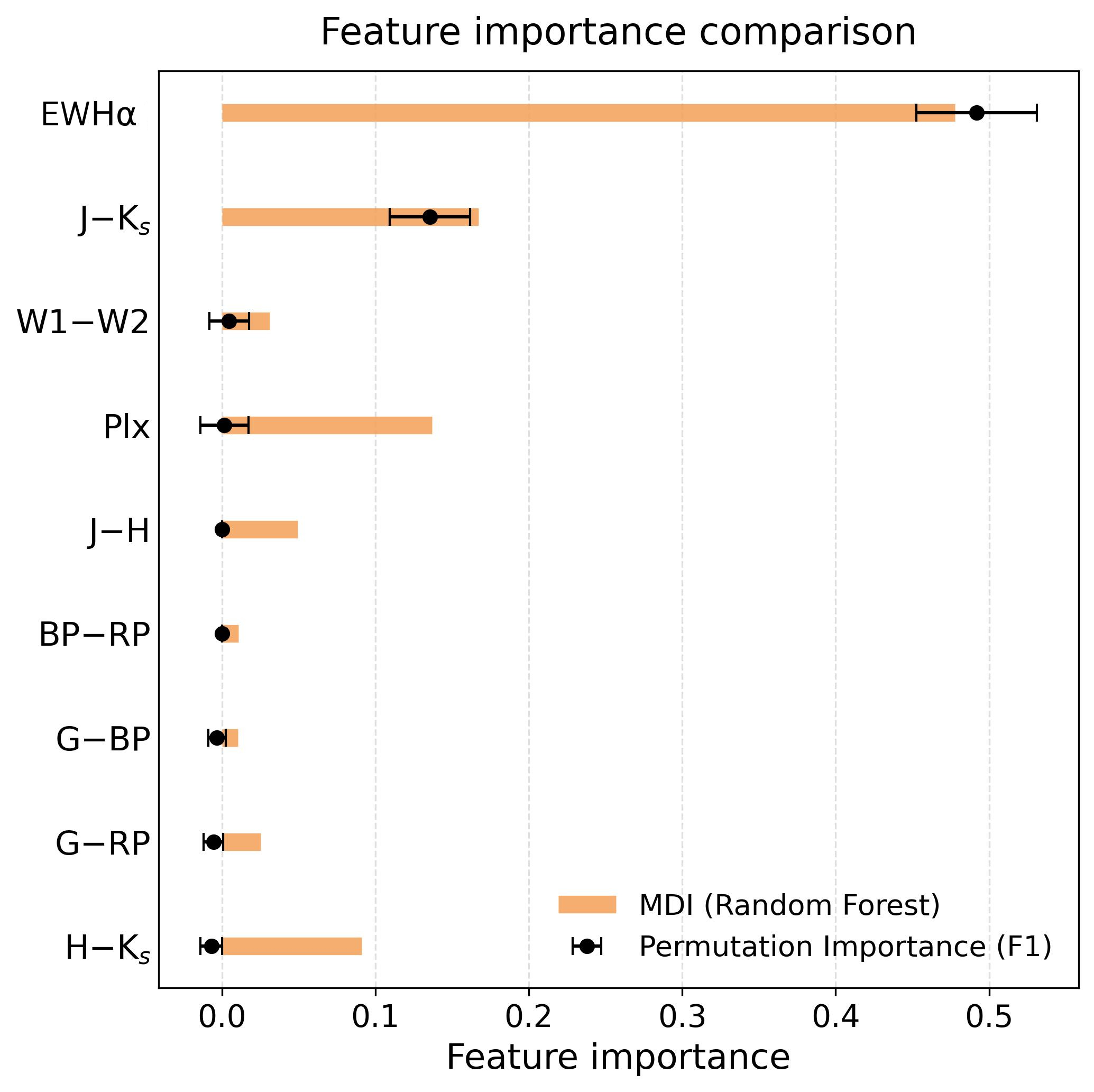}
    \caption{Comparison between the mean impurity decay of Random Forest (MDI, orange bars) and the importance of permutation features calculated using the F1 score (black dots).}
    \label{fig:featureimportance}
\end{figure}

To assess the relevance of the input features in the classification process, we evaluated feature importance using two complementary approaches: the Random Forest mean decrease in impurity (MDI) and permutation feature importance, the latter computed as the decrease in the F1-score after randomly shuffling the values of each feature. Figure~\ref{fig:featureimportance} presents a direct comparison between these two global feature-importance estimators.

Despite their different conceptual definitions, both methods consistently identify the EWH$\alpha$ as the most influential feature of the model. This result is entirely consistent with the spectroscopic properties of the S-type SySts used in this work, as approximately 95\% of the confirmed systems exhibit H$\alpha$ in emission. A detailed discussion of the role of H$\alpha$ is provided in Section 5.2.

After H$\alpha$, the J-K$_s$ near infrared colour emerges as the second most relevant feature in both importance estimators. This colour traces the photospheric emission of the cool giant component, whose spectral energy distribution peaks in the near-infrared (0.8-1.1 $\mu$m), as expected for late-type giants.

Beyond the J$-$K$_s$ colour, the relative importance of the remaining features is less clearly ordered. The two estimators no longer show a strict one-to-one correspondence in their ranking, and several parameters, including additional near-infrared colours, optical colour indices, and parallax, display comparable importance values. This behaviour indicates that, at this level, the classifier exploits a combination of correlated observables rather than relying on a single dominant parameter.

In particular, the contribution of parallax should not be interpreted as a direct physical diagnostic. In long-period binary systems such as S-type SySts, unresolved orbital motion can introduce systematic biases in Gaia astrometric solutions. While a parallax threshold was applied to ensure basic astrometric reliability, its role in the model likely reflects indirect correlations with distance-dependent photometric quantities rather than intrinsic physical information.

At the other end of the ranking, features with the lowest importance correspond to Gaia optical colour indices (G$-$BP, BP$-$RP). Owing to the extremely broad Gaia passbands, spanning from the near-UV to the near-infrared ($\sim$300–1,000~nm), these colours tend to dilute narrow but intense emission lines such as H$\beta$, [O,\textsc{iii}], and H$\alpha$, thereby reducing their discriminating power relative to infrared indices.

In addition, we have evaluated the performance of the model using a validation set composed of 12 confirmed S-type SySts taken from \citet{2024Lucy}. All of these stars are within the parameter range in which the model was trained, which ensures that the validation set is aligned with the characteristics of the dataset used in the training. As a result of this evaluation, the model was able to correctly recover 11 of the 12 SySts, implying a detection rate of 92.3\%. All recovered stars exceeded the previously proposed classification threshold of 70\%. The average probability of symbiotic class membership for these stars was 97\%, with a standard deviation of only 0.05\%, reflecting consistency in the model predictions. 

The only unrecovered source from the validation set is the variable star V V1918 Sgr*, a binary system previously classified in the literature as a planetary nebula, where the cool component has been identified as a K4~I supergiant. This classification is consistent with its extreme W1-W2 color, which placed the object edge of the the parameter space covered by out training sample. Although its photometric parameters formally lie within the model’s training domain, several of its features are sparsely represented in the training set. A local SHAP \citep{2017Shap} analysis of this source (Appendix~\ref{app:shap_waterfall}) indicates that its classification is influenced by a combination of photometric and spectroscopic features, including the EWH$\alpha$ , as well as the optical (G–BP) and infrared (W1–W2) colors. In particular, the H$\alpha$ values dominate the shift of the prediction below the symbiotic classification threshold, while W1–W2 and G–BP provides a secondary contribution to the final decision.

This case illustrates a limitation of the model when applied to objects with atypical stellar components and suggests that its performance may degrade for systems dominated by very luminous or otherwise peculiar donors, in which spectroscopic indicators such as H$\alpha$ may no longer play a dominant role, while broad-band optical and infrared colors become the primary drivers of the classification. These regimes are underrepresented in the training data.

To further evaluate the model$’$s robustness, we additionally tested the classifier using a validation set of 155 photometric mimics, of which 90 lie within our parameter space (i.e., misclassifications reported in the \citealt{2019Merc} catalogue). These sources were initially proposed as possible SySts but were later spectroscopically rejected. The test resulted in 51 objects being incorrectly classified as symbiotic, increasing the model’s contamination rate from 15.15\% to approximately 33.35\%. Among the contaminants, we identified 12 massive and supergiant stars, 10 planetary nebulae, 9 variable stars, 6 evolved giant stars, 5 young stellar objects, 4 subdwarfs, and 4 eclipsing binaries.

\section{Complementary analysis of SySts candidates}
\label{seccion5}

After validating the model's performance, we applied it to a filtered dataset of 2\,538\,939 objects, as described in Section~\ref{seccion2}, excluding the stars used in the training sets. This process yielded a total of 1\,559 candidate S-type SySts with probabilities higher than 50\%, from which we recovered 990 with probabilities exceeding 70\%. We then  performed a preliminary analysis of these sources using Gaia DR3 data (specifically the \texttt{mh\_gspphot}{} module \citep{2023Recio-Blanco}) and SkyMapper DR2 \citep{2019SkymapperDR2} photometry, with the aim of extracting key astrophysical and photometric parameters such as luminosity, \rm T$_{eff}$, [Fe/H], log g, and ${\rm EW}{\rm H\alpha}$. 

\begin{figure}[h!]
    \centering
    \includegraphics[width=1.0\linewidth]{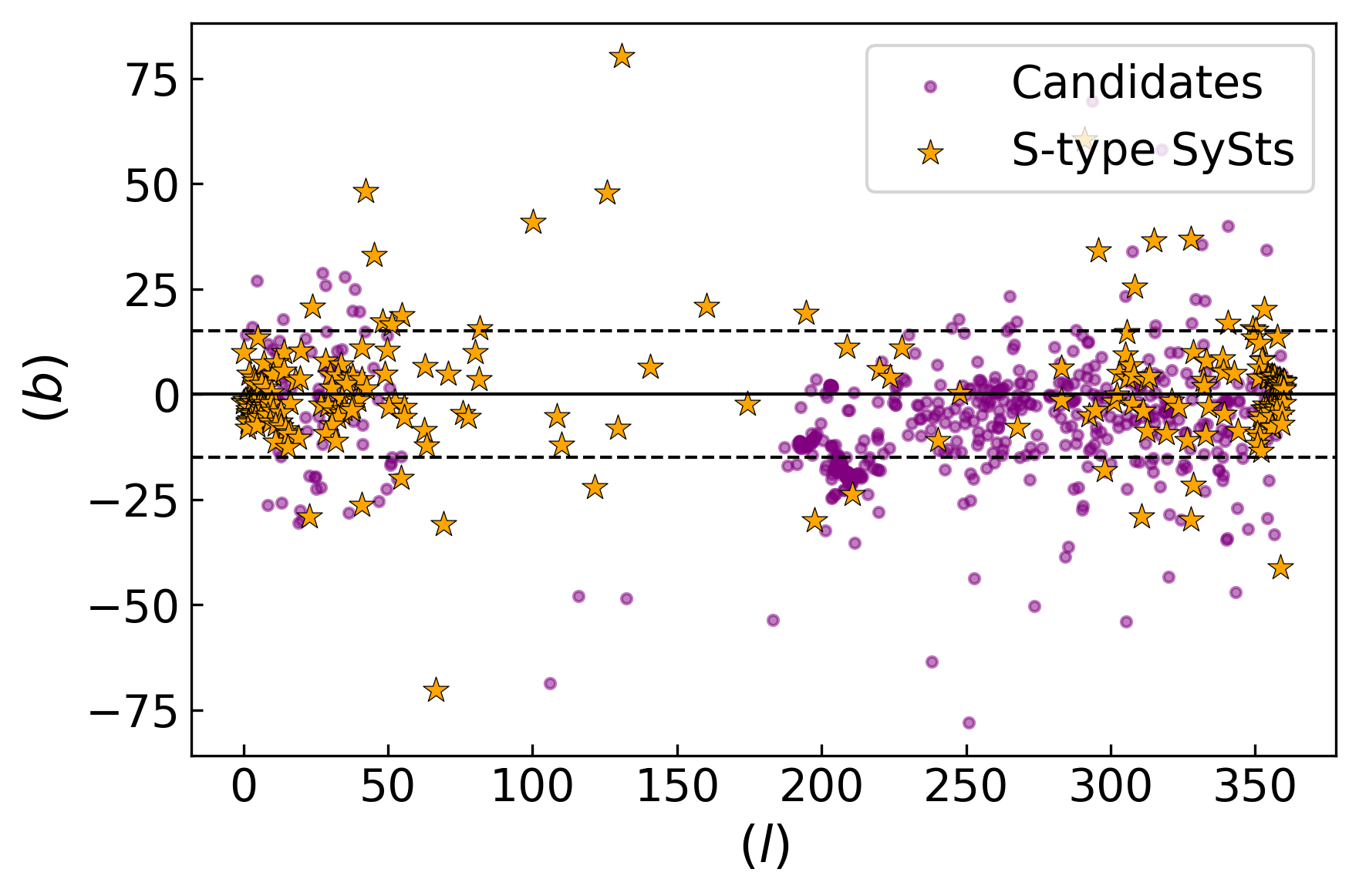}
    \caption{Galactic distribution of 990 SySts candidates (purple points) according to our model and confirmed S-type of SySts (orange stars).}
    \label{fig:galacticdistribution}
\end{figure}

The Galactic distribution of the confirmed S-type SySts and the candidates identified in this work is shown in Fig.~\ref{fig:galacticdistribution}, where the dashed lines mark the region $|b| < 15^\circ$, commonly associated with a higher density of SySts in the Milky Way \citep{2021Merc}. However, this apparent concentration toward the Galactic plane may be partly driven by selection effects, as large-scale surveys often focus on the disk. Previous studies, such as \citet{2021Munari}, have shown that SySts can also be found at higher Galactic latitudes, suggesting that the observed distribution may not fully reflect the intrinsic spatial population of these systems.

\subsection{Physical Parameters from Gaia}

Astrophysical parameters for the 990 candidates were retrieved using the Gaia DR3 \texttt{mh\_gspphot} module, subject to data availability. While these parameters offer valuable complementary diagnostics for candidate characterization, they must be interpreted cautiously. The atmospheric models used for parameter estimation \citep{2023Creevey} are not optimized for complex systems like SySts, which often exhibit variability and binary interaction effects.

Luminosities were estimated via absolute magnitude in the G band (\(M_G\)), calculated from apparent magnitudes, parallaxes, and extinction corrections. Extinction (\(A_G\)) was derived from reddening maps by \citet{1998SFD}, accessed through the \texttt{dustmaps} package \citep{sdf}. To assess evolutionary stages, candidates were classified based  \(M_G\): those with \(M_G < -0.5\) were considered likely AGB stars, while sources with \(-0.5 \leq M_G \leq 2.5\) were classified as RG. Approximately 67.3\% of objects with valid luminosities fall in the AGB regime and 8.6\% in the RGB regime, consistent with evolved giant stars.

The \(T_{\rm eff}\) mostly lie within 3\,500–4\,000~K, typical of M-type giants. However, about 15\% of candidates show higher values around 15\,000~K, which are too low to represent the hot white dwarf components (usually 50\,000–150\,000~K). These intermediate temperatures likely arise from composite spectra where the hot component, accretion disk, and nebular emission contribute to the optical flux, as observed in systems like T~CrB and CH~Cyg \citep{2015Zamanov,2018Stoyanov}. This complexity challenges the applicability of single-star atmospheric models.

The $[\mathrm{Fe}/\mathrm{H}]$ spans from $-4.1$ to $+0.8$, reflecting the chemical diversity expected across the Galaxy. Since SySts originate in varied galactic environments, this wide range, including solar and subsolar values, is consistent with confirmed samples, and thus not a strong membership discriminator.

The \(\log g\) values from Gaia range between -0.1 and 4.7, differing from the typical \(-1\) to 1 expected for S-type SySts \citep{2017Galan, 2023Galan}. Only 674 candidates fell within the expected range. This discrepancy likely stems from the intrinsic complexity of SySts, such as binarity, variability, and composite spectra limiting the reliability of Gaia’s single star models. Therefore, \(\log g\) should be interpreted cautiously and in conjunction with other parameters.

Importantly, none of these astrophysical parameters are used to exclude candidates. Sources lacking reliable Gaia estimates or falling outside expected ranges remain in the sample. Instead, this information helps prioritize targets for follow\-up, focusing on those whose physical properties most closely match confirmed SySts.

\subsection{$H{\alpha}$ Emission}

The values of ${\rm EW}{\rm H\alpha}$ in the candidates sample exhibit values ranging from $-$10.6 to $-$0.6~\AA, with a median around $-0.8$~\AA. This confirms the presence of H$\alpha$ emission in all sources, although generally weaker than in the confirmed sample of S-type SySts, which has a median ${\rm EW}{\rm H\alpha}$ of $-$4.2~\AA. This difference may reflect intrinsic variability or accretion-only states of the stars, or the inclusion of false positives, which according to the confusion matrix should occur in about 15\% to 33\% of the data.

Nevertheless, the ${\rm EW}{\rm H\alpha}$ values employed here correspond to the Gaia DR3 pseudo-equivalent widths (pEW), which are derived from low-resolution BP/RP spectra and rely on automated continuum estimation. As discussed by \citet{2023Creevey}, a temperature-dependent correction is applied for sources cooler than $T_{\rm eff} < 5000$~K, which may not be fully adequate for the cool giant components typical of symbiotic systems. Furthermore, \citet{2022Shridharan} found that Gaia’s pEW values can be systematically biased for emission-line stars due to the simplified background model adopted in the pipeline. These effects could lead to either an underestimation or overestimation of the actual emission strength, especially in composite or dust-affected spectra. Therefore, while the Gaia H$\alpha$ emerges as the most relevant feature in the Random Forest and permutation analyses (Section~\ref{seccion4}), its absolute values should be interpreted with caution, and future higher-resolution spectroscopy will be necessary to validate these measurements.

\subsection{SkyMapper Photometry}

\begin{figure}[h!]
\centering
\includegraphics[width=1.0\linewidth]{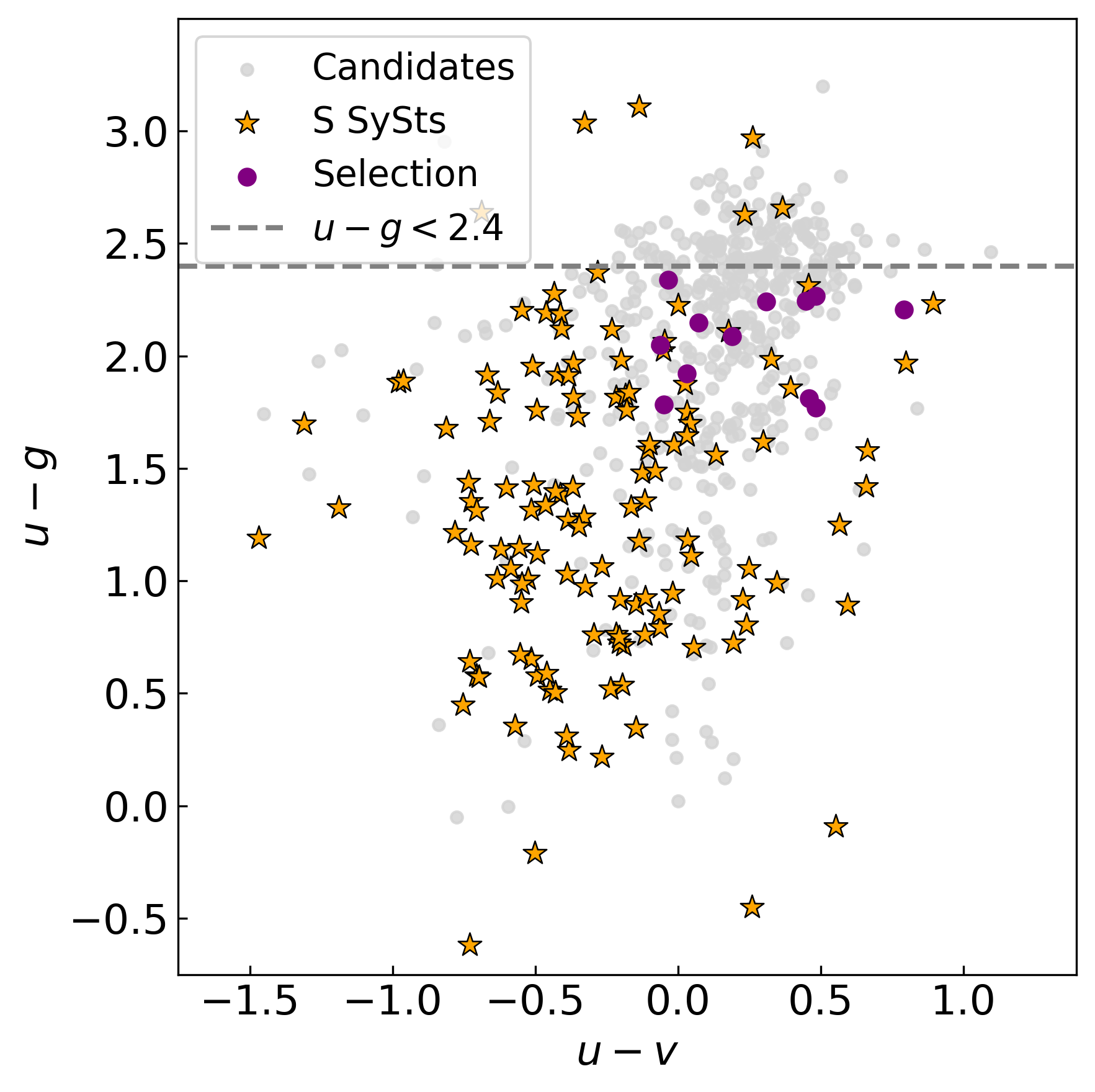}
\caption{Color-color plot of SkyMapper's $u-v$ versus $u-g$ photometry for our candidates (gray dots), known SySts (orange dots), and the 12 most likely candidates (purple dots).}
\label{fig:coloruv}
\end{figure}

We cross-compatibility tested our candidates with SkyMapper DR2 to study their photometric behavior in the ultraviolet regime. Of the 990 candidates, 660 presented valid $u$, $v$, and $g$ photometries, as shown in Fig.~\ref{fig:coloruv}. Following \citet{2024Lucy}'s criteria, we selected sources with $u < 16$~mag and $u - g < 2.4$, indicative of a hot, compact companion. A total of 145 candidates met these conditions.

Figure~\ref{fig:coloruv} presents the $u-v$ versus $u-g$ color-color plot, where the gray dots represent the full candidate sample and the orange symbol corresponds to the known SySts. These color indices effectively isolate symbiotic systems: $u-g$ is sensitive to the ultraviolet excess of the hot component, while $u-v$ provides an orthogonal axis that helps distinguish different stellar populations \citep{2024Lucy}. 

Most of the known SySts cluster in the lower left region, reflecting a strong ultraviolet excess, consistent with an active hot companion. A more dispersed group appears towards higher $u-g$ values, corresponding to systems with disk activity or nova-like eruptions, where the ultraviolet colors vary due to accretion changes or dimming. This color-space therefore serves as a powerful diagnostic tool for the photometric identification of SySts, capturing both classical hot systems and more complex or evolved objects with variable activity.

\subsection{Refined Candidate Selection}

To isolate the most promising candidates, we applied a multiparameter filtering approach based on the criteria summarized in Table~\ref{tab:selectioncriteria}. These included astrophysical parameters derived from \textit{Gaia} described above, such as \rm T$_{eff}$, M$_G$, log g, and [Fe/H], as well as constraints on ultraviolet excess indicators obtained using SkyMapper photometry.

The adopted thresholds were defined by combining two strategies: statistical cutoffs based on the distribution of confirmed S-type SySts, typically using the median value $\pm 3\sigma$, and empirical limits reported in the literature, particularly in the case of SkyMapper colors. This approach ensures that the selected subsample occupies a parameter space consistent with known interacting binaries, while allowing for physical diversity. After applying these filters to the full set of 990 high-confidence candidates, we identified a refined sample of 52 sources whose physical and photometric properties closely resemble those of confirmed S-type SySts. Among these, 12 objects also exhibit significant ultraviolet excess, as defined by our SkyMapper criteria, reinforcing their classification as likely systems interacting with a hot component. Table~\ref{tabla-datos} lists the main parameters of these 12 high-priority candidates, represented by purple dots in Fig.~\ref{fig:coloruv}.

It is important to emphasize that the strict selection applied here is intended to prioritize targets for immediate spectroscopic follow-up.

\begin{table*}[h!]
\centering
\caption{Selection criteria used to refine the candidate sample.}
\label{tab:selectioncriteria}
\begin{tabular}{lll}
\hline
Parameter & Accepted Range & {Selection Rationale} \\
\hline
$M_G$ & $-2.15$ to $6.07$ & 3$\sigma$ around the median of known S-type SySts. \\
$T_{\rm eff}$ & 2900 to 5700~K & 3$\sigma$ around the median peak of confirmed SySts. \\
Fe/H & $-2.49$ to $+0.80$ & Full range observed in known systems. \\
$\log g$ & 0.01 to 2.70 & Excludes main-sequence stars (see \citealt{2021Merc}). \\
SkyMapper Photometry & $u < 16$, $u - g < 2.4$ & From UV excess criterion in \citealt{2024Lucy}.
\\
\hline
\end{tabular}
\end{table*}

\begin{table*}[h!]
    \centering
    \caption{List of candidates for SySts based on Gaia DR3 parameters. Columns include right ascension (RA), declination (Dec), Gaia DR3 identifier, absolute magnitude ($M_G$), effective temperature ($T_\mathrm{eff}$), metallicity ([Fe/H]), surface gravity (logg), $H{\alpha}$ equivalent width (${\rm EW}{\rm H\alpha}$), and photometric magnitudes in the $u$ band and $u-g$ color index.}
    \begin{tabular}{ccccccccccc}
    \hline
    Ra & Dec & Gaia DR3 source ID & $M_G$ & $T_\mathrm{eff}$ & [Fe/H] & logg & ${\rm EW}{\rm H\alpha}$ & $u$ & $u-g$ \\
    \hline
  24.383 & -34.100 & Gaia DR3 5012312291397950848 & -0.526 & 3644 & 0.381 & 0.523 & -0.914 & 14.639 & 2.207\\
  93.997 & -13.712 & Gaia DR3 2993726727989383680 & -1.414 & 3921 & -0.048 & 0.485 & -0.671 & 14.769 & 2.337\\
  120.794 & -12.236 & Gaia DR3 5726316966180141440 & -1.093 & 3712 & 0.203 & 0.349 & -0.850 & 15.189 & 1.771\\
  198.601 & -59.782 & Gaia DR3 6061878912774021760 & -0.889 & 3718 & 0.387 & 0.620 & -0.773 & 15.880 & 2.086\\
  207.334 & -68.571 & Gaia DR3 5850238455010703744 & -0.556 & 5268 & -1.430 & 1.200 & -0.727 & 14.894 & 2.049\\
  262.301 & -14.841 & Gaia DR3 4137408314848139264 & -0.735 & 3841 & -0.042 & 0.417 & -0.764 & 15.364 & 2.148\\
  273.639 & -42.496 & Gaia DR3 6724544091084713600 & -1.094 & 3933 & -0.301 & 0.251 & -1.006 & 14.894 & 2.244\\
  274.889 & -36.077 & Gaia DR3 4038201133056103936 & -1.495 & 3754 & 0.254 & 0.521 & -0.754 & 14.890 & 2.265\\
  285.743 & -13.702 & Gaia DR3 4101579491517730816 & -1.055 & 3435 & 0.530 & 0.228 & -0.875 & 15.235 & 1.921\\
  292.062 & -74.657 & Gaia DR3 6367476693509172992 & -1.321 & 3907 & -0.163 & 0.116 & -1.270 & 15.134 & 1.811\\
  297.115 & 1.885 & Gaia DR3 4241327591186759808 & -1.136 & 3757 & -0.208 & 0.043 & -0.962 & 14.484 & 2.241\\
  300.322 & -15.045 & Gaia DR3 6877513268319887616 & -1.706 & 3667 & 0.286 & 0.086 & -0.773 & 13.841 & 1.785\\
    \hline
    \end{tabular}
    \label{tabla-datos}
\end{table*}

\section{Comparison with previous works}
\label{seccion6}

Our study focuses on the identification of new SySt candidates through machine learning techniques applied to public datasets. While previous works have employed similar approaches, our methodology stands out by utilizing the largest available catalog of point sources to date, along with the integration of low-resolution astrometric and spectroscopic data.

To validate and contextualize our classification results, we compared them against five previously published catalogs selected based on their similarity in feature sets used for training and classification. This comparison includes the total number of sources in each catalog, the subset within our model's defined parameter space, and the fraction of sources classified with probabilities exceeding 70\%. We also consider the original selection methods and present the kernel density distribution of classification probabilities for these samples.

Our findings reveal strong agreement with recent catalogs that employ photometric and spectroscopic diagnostics aligned with our feature space, particularly those emphasizing $H{\alpha}$ emission and consistent infrared properties. Meanwhile, observed differences with older or more heterogeneous samples highlight the importance of standardized parameter spaces and clearly defined selection criteria. Overall, these results support the effectiveness of trained and constrained machine learning classifiers as robust tools for the large-scale identification of symbiotic systems, especially shell-burning types characterized by strong $H{\alpha}$ emission and stable infrared signatures.

\begin{table*}[h!]
    \centering
    \small
    \caption{Comparison of our classification results with previously published catalogs of SySts candidates. The table shows the total number of sources in each catalog, the number within our model’s parameter space, the number and percentage of sources classified with probabilities above 70\%, and the original selection method used in each study.}
    \label{tab:comparison}
    \begin{tabular}{lrrrrl}
    \hline
    Reference & Total & Parameter Space & Classified & \% of $>$70\% & Original Selection Method \\
    \hline
    \citet{2008Corradi}  & 1068 & 465 & 456 & 42.7\% & Narrow-band photometry H$\alpha$ + 2MASS IR \\
    \citet{2019Akras}    & 410  & 150 & 125 & 30.4\% & SEDs from 2MASS + WISE + Gaia DR2 temperature estimates \\
    \citet{2022Akras}   & 125  & 64  & 46  & 36.8\% & Decision tree ML using near- and mid-IR color indices \\
    \citet{2019Merc}     & 730  & 259 & 109 & 14.9\% & Literature compilation of candidates \\
    \citet{2024Lucy}    & 10   & 10  & 7   & 70.0\% & SkyMapper $uvg$ photometry \\
    \hline
    \end{tabular}
\end{table*}

\begin{figure}[h!]
    \centering
    \includegraphics[width=1.0\linewidth]{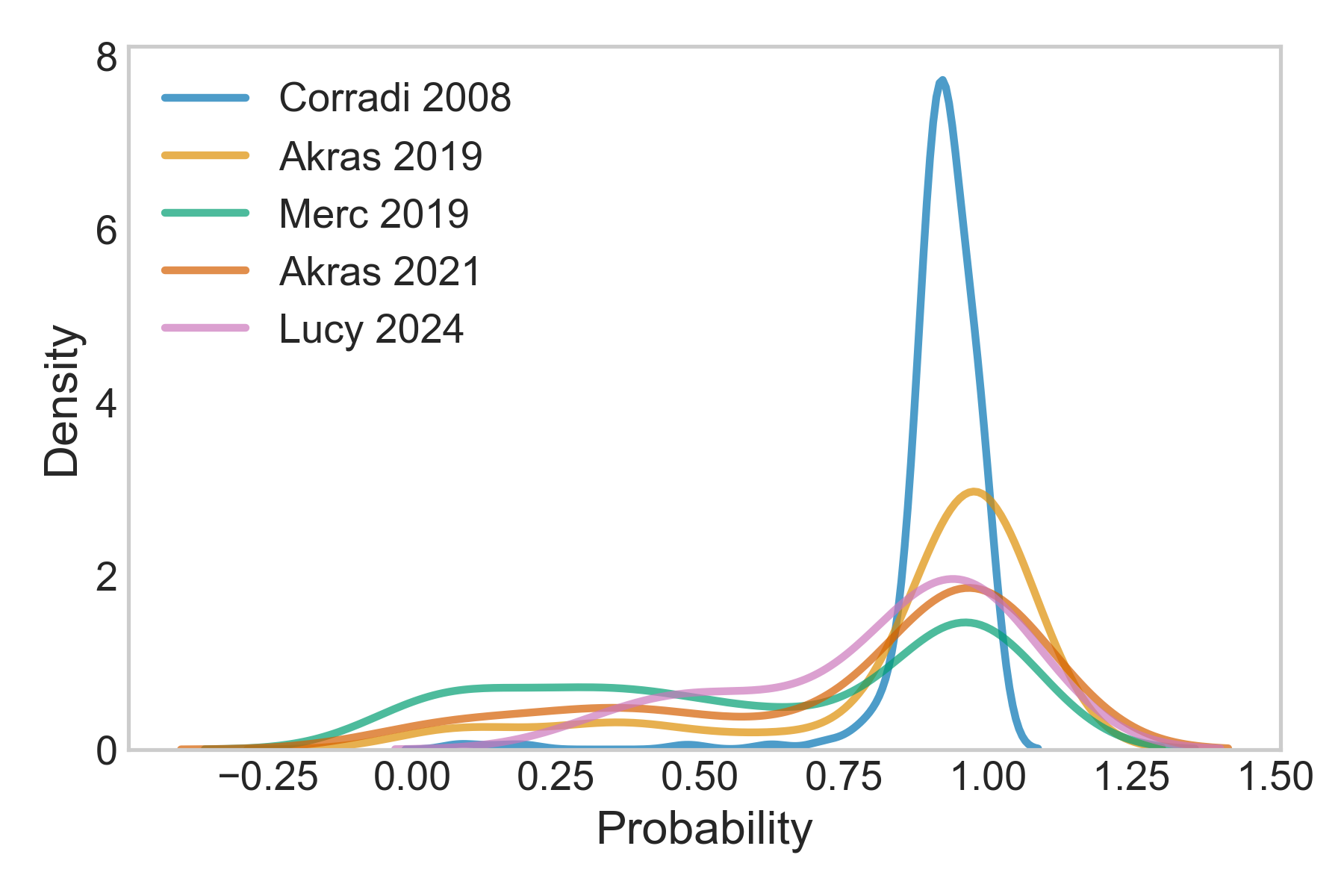}
    \caption{Kernel density distribution of the classification probabilities assigned by our model to SySt candidates proposed in previous works. The Merc$'$s catalog includes literature-based candidates compiled between 2019 and November 2024.}
    \label{fig:probabilities}
\end{figure}

\section{Summary and conclusions}
\label{seccion7}

In this work, we present a new supervised classification methodology for identifying S-type SySts candidates, based on an RF algorithm combined with SMOTE to address class imbalance. The model is trained with a set of astrometric, photometric, and spectroscopic features derived from Gaia DR3, 2MASS, and WISE data. Key input variables include color indices, parallax, and the ${\rm EW}{\rm H\alpha}$ line. These features define a restricted parameter space where symbiotic systems are most likely to reside, enabling efficient candidate selection.

The classifier demonstrates excellent overall performance, especially in distinguishing non-SySt sources. While classifying the minority class (S-type SySt) remains more complex, the use of SMOTE helped mitigate the effects of class imbalance. To refine candidate selection, we applied a classification probability threshold of 70\%, selected to balance completeness and accuracy across the entire candidate sample.

By applying our model to a selected sample of over 2.5 million sources, we identified 990 high-probability candidates. Among them, 12 exhibit physical properties consistent with known S-type SySts, including H$\alpha$ emission and ultraviolet excess in the SkyMapper $uvg$ bands. In addition, we identified a group of 133 candidates with similar photometric characteristics (i.e., blue optical colors and infrared excess). While their classification probabilities are high, the lack of complete astrophysical information (such as \rm T$_{eff}$ or log g) prevents a more detailed evaluation in this work. Nevertheless, these sources represent strong candidates for future observations, as they could host hot compact companions and constitute previously unknown interacting binary systems.

The model's performance was independently verified with a validation set introduced by \citet{2024Lucy}, which correctly recognized 11 of the 12 known S-type SySts ($\sim$92.3\%), demonstrating robust predictive capability. Importantly, the model was trained and validated on a well-characterized dataset with broad coverage of the spectral energy distribution (SED), including spectroscopic features consistent with data availability in the Merc$'$s catalog. Applying the model to a larger dataset, or one with coarse SED sampling, could reduce its predictive reliability. This is because the classifier cannot accurately recognize objects outside the parameter space it was trained on. To further assess its robustness, we tested it on a sample of 155 photometric mimics reported by \citet{2019Merc}, finding that 51 were incorrectly classified as symbiotic, which raises the contamination rate to $\sim$33\%. These results confirm that the classifier performs reliably within its trained parameter space but is sensitive to sources with incomplete or coarsely sampled SEDs.

We also applied our model to previously published candidate catalogs, observing an overall agreement close to 50\%, with greater consistency in samples incorporating H${\alpha}$ and infrared criteria. This agreement validates the robustness of our classifier. This bias could lead to underrepresentation of accretion-only systems such as SU~Lyn, whose emission lines are not as prominent compared to shell-burning systems. Furthermore, low-resolution spectra can underestimate H${\alpha}$ intensity due to continuum contamination from the red giant, highlighting the need for caution in emission-based selection.

The limited number of confirmed SySts, even when complemented by our new candidates, has significant implications for population synthesis models. While adding several hundred confirmed systems would establish meaningful lower bounds, our current census remains well below theoretical predictions, which reach up to $(53 \pm 6) \times 10^{4}$ systems. This discrepancy may reflect observational biases, particularly against accretion-only systems with weak emission lines, or highlight incomplete evolutionary modeling. Addressing this gap is essential to advance our understanding of binary evolution and the galactic symbiotic population.

Although infrared emission and Balmer emission lines are classic indicators of SySts, they are not exclusive diagnostic. Other astrophysical sources can mimic these features, emphasizing the importance of a multiwavelength approach that integrates photometric and spectroscopic data to reduce contamination and improve classification reliability for appropriate spectroscopic follow-up. In this context, we plan to retrain the model with new S-PLUS data, which offers broader SED coverage and could enhance classification performance. These observations are currently in progress, and the updated model will be presented in future work.

In general, this study demonstrates that MLs are powerful tools for the automated discovery of rare astrophysical systems in large-scale public databases. When trained with carefully curated representative datasets, classifiers can effectively model complex relationships between observables and generate high-confidence predictions. For SySts, the limited number of confirmed examples and the diversity of spectral manifestations pose significant challenges. Therefore, results must be interpreted in an astrophysical context and validated through rigorous observational follow-up.

\begin{acknowledgements}

We thank the referee for their constructive comments and valuable suggestions, which helped to improve the clarity and quality of this manuscript.
This work was supported by DIDULS/ULS through projects PTE23538510 and PTE23538516, as well as by the ANID FONDECYT project 1231637.
M.J.A. acknowledge support from the ANID FONDECYT Iniciación 11251912. NEN and GJML are members of the CIC-CONICET (Argentina).
This work presents results from the European Space Agency (ESA) space mission Gaia. Gaia data are being processed by the Gaia Data Processing and Analysis Consortium (DPAC). Funding for the DPAC is provided by national institutions, in particular the institutions participating in the Gaia MultiLateral Agreement (MLA). The Gaia mission website is https://www.cosmos.esa.int/gaia. The Gaia archive website is https://archives.esac.esa.int/gaia.
This research has made use of the VizieR catalogue access tool, CDS, Strasbourg, France, as well as public data from the Two Micron All Sky Survey (2MASS) and the Wide-field Infrared Survey Explorer (WISE). 2MASS is a joint project of the University of Massachusetts and the Infrared Processing and Analysis Center/California Institute of Technology, funded by NASA and the NSF. WISE is a joint project of the University of California, Los Angeles, and the Jet Propulsion Laboratory/California Institute of Technology, funded by NASA.
This work has also made use of data from the SkyMapper Southern Survey. The national facility capability for SkyMapper has been funded through ARC LIEF grant LE130100104 from the Australian Research Council, awarded to the University of Sydney, the Australian National University, Swinburne University of Technology, the University of Queensland, the University of Western Australia, the University of Melbourne, Curtin University of Technology, Monash University, and the Australian Astronomical Observatory. SkyMapper is owned and operated by the Australian National University’s Research School of Astronomy and Astrophysics. The survey data were processed and provided by the SkyMapper Team at ANU. The SkyMapper node of the All-Sky Virtual Observatory (ASVO) is hosted at the National Computational Infrastructure (NCI). Development and support of the SkyMapper node of the ASVO has been funded in part by Astronomy Australia Limited (AAL) and the Australian Government through the Commonwealth's Education Investment Fund (EIF) and National Collaborative Research Infrastructure Strategy (NCRIS), particularly the National eResearch Collaboration Tools and Resources (NeCTAR) and the Australian National Data Service Projects (ANDS).
      
\end{acknowledgements}

\bibliographystyle{aa}
\bibliography{referencias.bib}

\begin{appendix}

\onecolumn
\section{ADQL query used for the initial sample extraction}
\label{app:adql}

The following ADQL query was used to extract the initial candidate sample from Gaia DR3, cross-matched with 2MASS and including astrophysical parameters. The selection criteria were defined to match the parameter space covered by the confirmed symbiotic stars used for training, ensuring consistency between the training and application domains.

\begin{verbatim}
SELECT 
    gaia.source_id, gaia.*, ap.*, tmass.*
FROM gaiadr3.gaia_source AS gaia
JOIN gaiadr3.tmass_psc_xsc_best_neighbour AS xmatch USING (source_id)
JOIN gaiadr3.tmass_psc_xsc_join AS xjoin USING (clean_tmass_psc_xsc_oid)
JOIN gaiadr1.tmass_original_valid AS tmass 
    ON xjoin.original_psc_source_id = tmass.designation
JOIN gaiadr3.astrophysical_parameters AS ap 
    USING (source_id)
WHERE
    gaia.phot_g_mean_mag - gaia.phot_rp_mean_mag BETWEEN 0.29 AND 2.34
    AND gaia.phot_g_mean_mag - gaia.phot_bp_mean_mag BETWEEN -4.19 AND 0.10
    AND gaia.phot_bp_mean_mag - gaia.phot_rp_mean_mag BETWEEN 0.22 AND 5.95
    AND gaia.phot_g_mean_mag < 16
    AND gaia.parallax IS NOT NULL
    AND gaia.parallax_over_error > 10.00
    AND ap.ew_espels_halpha IS NOT NULL
    AND ap.ew_espels_halpha BETWEEN -18.49 AND 0.69
    AND tmass.j_m - tmass.h_m BETWEEN 0.22 AND 2.99
    AND tmass.j_m - tmass.ks_m BETWEEN 0.46 AND 5.33
    AND tmass.h_m - tmass.ks_m BETWEEN 0.11 AND 2.34
\end{verbatim}

\section{Kolmogorov--Smirnov tests for negative-class subsampling}
\label{app:ks_tests}

\begin{table}[h!]
\centering
\caption{Two-sample KS test comparing the selected negative subset (1\,600 sources) with the full negative population ($\sim$150\,000). All features have $p>0.05$.}
\label{tab:ks_pvalues}
\small
\setlength{\tabcolsep}{4pt}
\begin{tabular}{lcc lcc lcc}
\hline
Feature & KS & $p$ &
Feature & KS & $p$ &
Feature & KS & $p$ \\
\hline
Parallax  & 0.0231 & 0.5005 &
$BP-RP$   & 0.0161 & 0.8917 &
$G-BP$    & 0.0139 & 0.9645 \\

$G-RP$    & 0.0168 & 0.8581 &
$J-H$     & 0.0157 & 0.9061 &
$J-K_s$   & 0.0200 & 0.6799 \\

$H-K_s$   & 0.0211 & 0.6125 &
$W1-W2$   & 0.0173 & 0.8369 &
${\rm EWH}\alpha$ & 0.0139 & 0.9629 \\
\hline
\end{tabular}
\end{table}

\section{Local SHAP explanation for a misclassified validation source}
\label{app:shap_waterfall}

\begin{figure}[h!]
    \centering
    \includegraphics[width=0.46\linewidth]{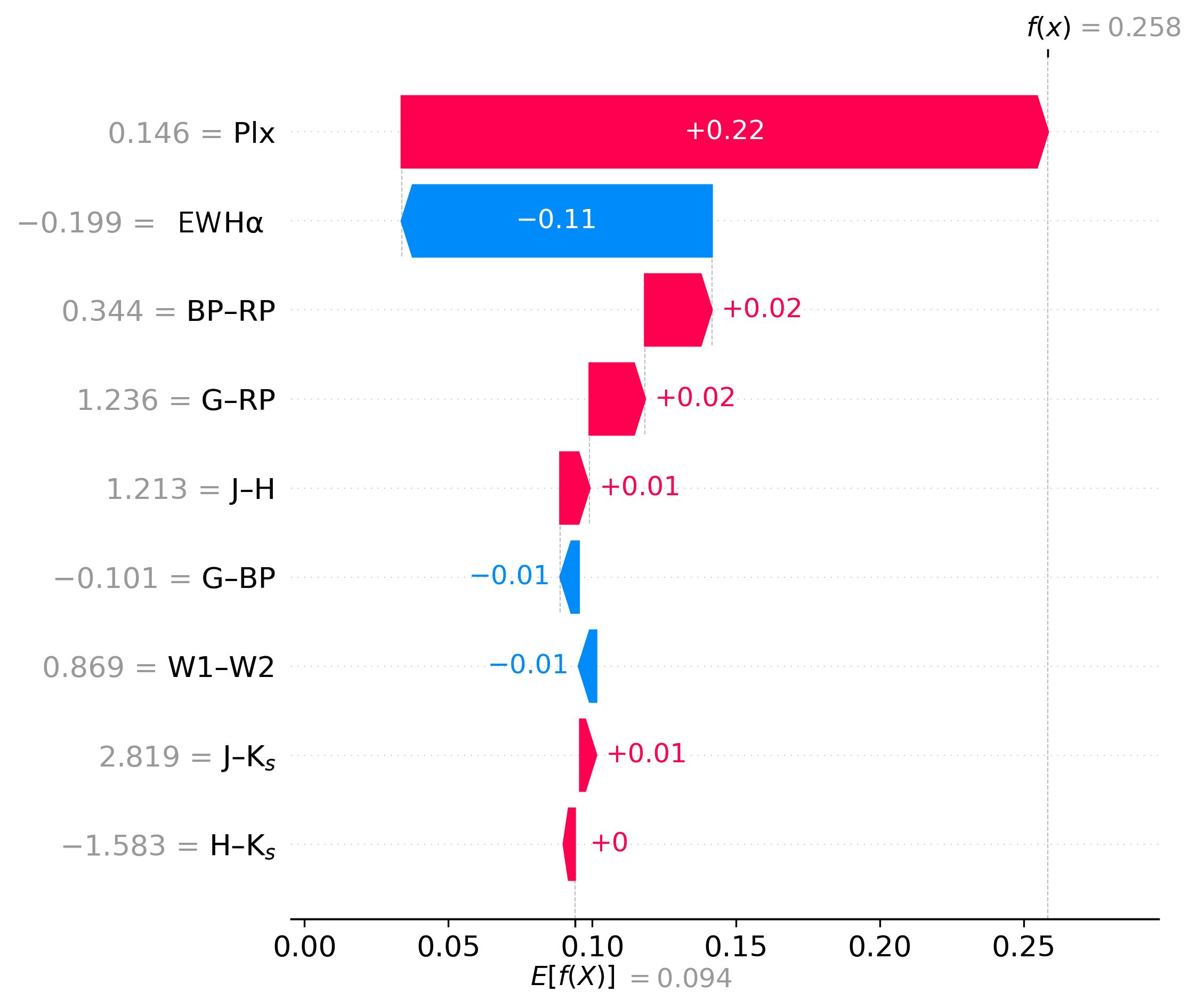}
    \caption{SHAP waterfall plot showing the local explanation of the classifier prediction for the validation source V~V1918~Sgr*. Positive contributions increase the predicted probability of the symbiotic class relative to the baseline.}
    \label{fig:shap_waterfall_vv1918}
\end{figure}

\end{appendix}

\end{document}